\documentclass[preprintnumbers,floatfix,twocolumn,aps,prl,unsortedaddress,superscriptaddress,longbibliography,showkeys]{revtex4-2}

\usepackage[para]{threeparttable} 
\usepackage{amsmath}
\usepackage{graphicx} 
\usepackage{booktabs}
\usepackage{subcaption}
\usepackage{url}
\usepackage{hyperref}
\usepackage{miller}
\usepackage{bm}
\usepackage{enumitem}
\usepackage{color}
\usepackage{siunitx}
\usepackage[utf8]{inputenc}
\begin{document}

\title{Machine-learned interatomic potential for sputtering of tungsten-boron surfaces}

\author{A. Bergero}
\affiliation{Department of Physics, P.O. Box 43, FI-00014 University of Helsinki, Finland}
\author{J. Byggmästar}
\affiliation{Department of Physics, P.O. Box 43, FI-00014 University of Helsinki, Finland}
\author{F. Granberg}
\thanks{Corresponding author}
\email{fredric.granberg@helsinki.fi}
\affiliation{Department of Physics, P.O. Box 43, FI-00014 University of Helsinki, Finland}

\begin{abstract}

Boronization, where boron is deposited onto tungsten surfaces, is a key technique to reduce plasma contamination, such as oxygen in Tokamak fusion reactors. The exact interaction between the boron atoms and the tungsten surface, and the effect of the harsh environment on these surfaces are however not fully understood, partially due to the lack of accurate atomistic simulations and interatomic potentials. Here, we develop a machine-learned interatomic potential for sputtering studies of W-B structures and deposition of boron onto tungsten surfaces. The machine-learned potential is trained to density functional theory data and allows accurate large-scale molecular dynamics simulations. Our aim is to understand how boron behaves when deposited on tungsten and how tungsten and boron are sputtered under irradiation. We observe that both the surface configuration/orientation and the surface composition affect the sputtering, and that depositing boron onto tungsten surfaces produces a dense boron layer. The developed potential shows good accuracy for both surface and bulk properties and can be used for simulations of mixed W and B systems.

\end{abstract}


\maketitle

\section{Introduction}

The constant increase in energy consumption, driven by habits, technological improvement, and globalization, requires a large-scale, low-carbon source of energy in the future. Fusion energy is a very promising candidate fitting all the aforementioned points. Additionally, it reduces the possible unstable supply chain or reliance on geopolitics, supporting long-term independence. Although research has been conducted for decades, especially in the Tokamak configuration \cite{introTOK}, there are still phenomena that remain poorly understood. One of the critical challenges is the plasma facing materials (PFMs), which must withstand high-dose irradiation, high temperatures, and endure various forms of damage and progressive changes resulting from extreme conditions, such as impacts by charged particles and/or neutrons \cite{introplasma1}. In the last decades, many experimental and computational studies have been performed to understand radiation damage starting from primary damage \cite{review-raddamge}. Computational studies on plasma-wall interaction have emerged and become more popular in recent decades due to the increased capabilities of high-performance computing infrastructure \cite{review-modeling}.

Currently, tungsten is considered the primary candidate for use as the plasma-facing material in ITER \cite{introITER}, and most likely in DEMO, specifically as the divertor and main chamber walls \cite{introITER,introTOK}. Tungsten sputtering has been computationally investigated for different incoming ions, such as D \cite{sputD,expW3ArD}, He \cite{sputOrientation}, Ar \cite{sputOrientation,sputA,peak,expW3ArD}, and W \cite{sputGene1,expW1Ar}. The sputtering dependence on the surface orientation of tungsten has also been investigated \cite{sputOrientation}. The sputtering yield has been observed to be influenced by many factors, such as impact energy, incoming ion angle, surface orientation and sometimes on surface temperature \cite{peak,sputOrientation,sputEro}. Experimentally, tungsten has been studied for a long time and sputtering yield by various particles, such as Ar \cite{expW1Ar}, W \cite{HECHTL90} and D \cite{expW3ArD}, can be found in the literature.

ITER decided to change the wall material from beryllium to tungsten due to toxicity and high erosion rate making it not suitable for the fusion roadmap \cite{WONGboro}. In addition, beryllium has a high sputtering yield, which would introduce a lot of impurities in the plasma. Sputtering is an undesirable effect, as a small concentration of heavy impurities can be more damaging than a large quantity of light ones, but if the amount of light impurities is too large, the benefits of having light elements are lost. The perfect candidate would be a material with low mass and low sputtering yield. A balance has to be found to minimise wall erosion and plasma cooling due to impurities. With pure tungsten, a new wall conditioning system had to be chosen that captures oxygen impurities, as free oxygen present in the reactor would increase radiative losses \cite{ITERboro2}. Boronization was chosen, where a thin layer of 10--100 nm of boron is deposited on top of the tungsten surfaces \cite{ITERboro}. Some studies suggest that a single boron application could be effective in capturing oxygen for anywhere between 2.5 and 12.5 weeks of campaign time \cite{Bcampagne}, after which, the boron oxide either saturates with oxygen, or erodes due to sputtering. The DEMO reactor is planned to have diborane to act as a pumping wall, reducing hydrogen recycling, oxygen impurities, and thus improving plasma performance \cite{EUROfusion2024_boronise_tokamak}. 

Experimental studies of boron self-sputtering have been conducted in the past decades \cite{YAMAMURA95} which shows a experimental value higher than the computational calculations at normal incident angle \cite{HECHTL92}. In addition, there are some computational studies for boron impacted by Ar, such as Stopping Range of ion in Matter (SRIM) simulations \cite{MahneSRIMsputB} showing that sputtering yield is highly dependent on the incoming angle of the ion for boron. A neural network potential (NNP) used with molecular dynamics (MD) for the sputtering of boron and boron-oxide by B and D incoming particles \cite{NNP25} shows higher sputtering yield for boron at 60 degrees incoming angle for deuterium ions. Boron impacted by D suggests that chemical sputtering could be a significant factor in limiting the boron coating lifetime, when exposed to low ion energies \cite{reacB}. Boron deposition in a small quantities has been studied using Monte Carlo methods \cite{DORFMANdepoB}, showing that the conditions of the metalloid adhesion are influenced by the directional bonding nature and the structural reconstruction of the substrate surface. The impact of thickness and uniformity of the boron layer deposited by the boronization process has been experimentally studied \cite{WUdepoBexp}, highlighting an experimental method able to diagnose the boron layer. More detailed insight about the boronization of tungsten and subsequent sputtering is needed from MD simulations. At the moment, however, there is no interatomic potential for simulations of sputtering from W$_x$B$_y$ surfaces.

Machine learning (ML) has emerged as a central tool in atomistic modelling, particularly for developing accurate interatomic potentials with capabilities beyond conventional analytic interatomic potentials \cite{introMLmuel,introMLmish}. ML potentials can achieve an accuracy close to that of the underlying quantum-mechanical methods used to generate training data, while bypassing explicit electronic-structure calculations \cite{QuantAccBart}, thus offering a great compromise between predictive fidelity and computational expense \cite{introMLderi}. An increasingly diverse ecosystem of ML interatomic potential frameworks has been developed, distinguished primarily by their representations of local atomic environments (descriptors) and choice of ML regression architecture \cite{introMLnnp,QuantAccBart}. Different ML potentials show different trade-offs between accuracy and computational speed \cite{introCOSTzuo}. For simulations requiring large systems or many simulations for statistics, like sputtering, a good balance between speed and accuracy becomes the most critical consideration. The tabulated Gaussian Approximation Potential framework (tabGAP) \cite{JesperHEA,QuantAccBart} was developed with this in mind and has been successfully used for radiation damage and sputtering simulations in fusion-relevant materials~\cite{JesperHEA,jespDESCR,AlexandreHEA}. 

In this work, we develop the first interatomic potential for W-B surfaces, using the tabGAP framework. The ML potential is aimed for large-scale or high-throughput sputtering and deposition simulations. After a detailed validation and discussion, we use the ML potential to simulate sputtering of W, B, and mixed W-B surfaces as well as investigate boronization of tungsten surfaces.

\section{Methods}

\subsection{Potential details}

The GAP model was trained using the QUIP package \cite{QuantAccBart} and subsequently tabulated into a tabGAP \cite{JesperHEA}. Tabulation requires training with only low-dimensional descriptors for the local atomic environments. We used a combination of a two-body interatomic distance descriptor, a three-body angular descriptor, and an Embedded Atom Model (EAM) density descriptor \cite{JesperHEA}. An analytical short-range baseline interaction was included as an external screened Coulomb pair potential following the methods in Ref. \citenum{jesperW}. The regularisation uncertainties for training were set to 0.001, 0.05 and 0.6, respectively, representing the energies (eV/atom), forces (eV/Å), and virials (eV). For the two-body, the cut-off was set to 5 Å and a transition width of 1 Å, using a squared-exponential covariance, the kernel amplitude parameter $\delta$ = 10.0, and 20 sparse environments uniformly sampled. The three-body angular descriptor has a cut-off of 4.1 Å and a transition width of 0.7 Å, a kernel amplitude parameter $\delta$ = 0.1, and 300 uniformly selected sparse points. The EAM-density descriptor using a generalised Finnis-Sinclair density function of third-order \cite{JesperHEA}. The cut-off radius was set to 5 Å, the kernel amplitude parameter $\delta$ = 1.0, and 20 uniformly selected sparse points. For more details on the theory behind the Gaussian Approximation Potential framework, refer to Refs. \citenum{gapbartok} and \citenum{RasmussenWilliams2006}. The input and output files from the training are available from Ref.~\cite{ZENODO}.

\subsection{Density functional theory}

The DFT training structures were calculated using the Vienna Ab initio Simulation Package (VASP) \cite{vasp1,vasp2,vasp3,vasp4}. The exchange–correlation energy was described using the generalised gradient approximation (GGA) in the Perdew–Burke–Ernzerhof (PBE) parametrization \cite{PBE}. The projector-augmented wave (PAW) method was used to describe the interaction between ions and electrons \cite{PAW1,PAW2} with the recommended PBE PAW datasets in VASP (W\_sv and standard B). The plane wave cut-off energy was 500 eV. Brillouin zone sampling used Monkhorst–Pack grids \cite{kpoint}, with consistent density controlled by a KSPACING of 0.15 Å$^{-1}$. Electronic occupations were smeared using Gaussian smearing with a width of 0.05 eV. The convergence in forces and energy was set to EDIFF = 10$^{-6}$ eV and the precision to ``accurate''.

\subsection{Molecular statics and dynamics simulations}

All classical molecular dynamics simulations were performed using the Large-scale Atomic/Molecular Massively Parallel Simulator (LAMMPS) \cite{lammps} with a maximum time step of 0.5 fs for B and WB, and 2 fs for W. LAMMPS was used both during training of the potential to obtain initial guesses for representative structures and later for the full sputtering and depositions simulations. Molecular statics calculations such as validations were done within the atomic simulation environment (ASE) framework \cite{ASE}. The details for the sputtering /depositions MD simulations are described in Section APPLICATIONS/Methods. 

\subsection{Training}

The accuracy of the tabGAP depends on the size and quality of the training data set. The database is broken down into four groups. The first group is pure W that was taken from previous work \cite{jesperW}. The second group we call "bulk", which contains bulk B and W-B structures with lattice distortion or thermal displacements. The third is for liquids and the fourth and last group are all configurations directly related to surface physics such as deposition, adsorption sites, surface vacancies, and dimers. More details can be found for the WB part in the supplementary Table I, and for pure W in Table II in Ref. \cite{jesperW}. The training data and final potential files are openly available from Ref.~\cite{ZENODO}.

Building on the pure W database~\cite{jesperW}, we started by adding isolated atoms and dimers B$_2$ and WB from 30 static calculations with interatomic distances from 0.5 Å (for B-B) and 0.8 Å (for W-B) to 5 Å. The dimers ensure accurate repulsion and bond energies between all element pairs. The minimum allowed dimer distance in VASP was determined by comparison to all-electron DFT data \cite{nordlund_repulsive_2025a}. To reproduce realistic repulsion down to zero distance, we also include pre-fitted and fixed screened Coulomb pair potentials following the same methods as previously for W-W \cite{jesperW}.

For sampling bulk phases we produced isotropic elastic deformations to all bulk B and W-B structures from the Materials Project \cite{matproj} (B: mp-160, mp-161, mp-1193675, mp-22046, mp-570316, mp-1228790, mp-1196985, mp-1104251, mp-570602, mp-632401, mp-1202723, mp-1182425, mp-1055985, mp-541848, mp-729184, and for W-B: mp-569803, mp-7832, mp-1113, mp-1008487, mp-10144, mp-1001602, mp-8079, mp-570938, mp-29651, mp-1205418). A factor of 0.9 to 1.1 was applied to each dimension to obtain 10 different elastic deformations. For sampling of finite-temperature atomic displacements, we choose only what we define as the relevant phases, mp-160, mp-161, mp-1193675, mp-1008487, mp-7832, mp-569803 and mp-1113, which are respectively the $\alpha$-trigonal, $\beta$-trigonal, $\alpha$-orthogonal, WB-orthogonal, WB-tetragonal, WB$_2$, and W$_2$B phases. Frames sampled from thermal displacement simulations at 300 K, 600 K and 1000 K were iteratively added to the training data. The first iteration was generated by ab-initio MD using VASP for $\alpha$-trigonal and WB-tetragonal phases. This was done to get an initial version of the tabGAP that is at least stable under finite-temperature MD. A second iteration was done using this trained potential including all above-mentioned relevant phases as well as mp-729184, named $\gamma$-plutonium in this work.

Liquids and some amorphous phases were iteratively introduced by melting cells of the relevant phases with initial versions of the potential in simulation from 300 K to 5000 K, and then cooling it down back to 300 K over a total simulation time of 1 ns. Ten frames were sparsely selected for each relevant phase considering all stages of heating, melting, possible recrystallization, and cooling. Two additional training iterations were done to make sure that the potential has iteratively learned accurate melting and liquid properties.

Surface training data were added by taking five relaxed frames for all relevant phases similar to the thermal displacement data, but with vacuum to make open surfaces. A first iteration was created with an initial potential for the boron structures, while for W$_x$B$_y$ structures we used ab-initio MD since the initial potential did not predict stable surfaces. An additional iteration was added for W$_x$B$_y$ using the new potential after the training from the first iteration. Additional training data were added by slicing the W$_x$B$_y$ surface configurations in order to have different atomic configurations at the surface.

A first iteration of surface vacancies on tungsten and some boron deposited on tungsten was done by sampling from ab-initio MD. Boron was randomly deposited on the surface at very low energy for both the \hkl(100) and \hkl(110) surfaces. The following iteration was done using the updated potential for the adsorption site Hollow four-fold (4F) on the W \hkl(100) surface, and the adsorption sites Hollow three-fold (3F), long bridge (LB), short bridge (SB), and top (TOP) for the W \hkl(110) surface \cite{adsorp}. A last iteration of the vacancies was performed including the adsorption sites for the \hkl(111) surface Bridge-Long (BL$_{32}$), that is a variation of LB but a deeper between the atomic row number 2 and 3, Faced-Center-Cubic hollow (FCC), and TOP \cite{adsorp}, and for the \hkl(112) surface 3F, SB, and TOP \cite{adsW}.

Boron deposited with 5 eV kinetic energies on tungsten cells thermalised to 300 K was part of many of the training iterations, with depositions ranging from a few atoms to full atomic layers. The first deposition was included with surface vacancy training data with 1 -- 2 randomly deposited boron atoms on tungsten \hkl(100) and \hkl(110) surfaces. A second deposition was included at the same time as the adsorption sites and surface training data by taking seven frames of a single layer of deposited boron on tungsten with surface orientation \hkl(100). A third deposition was introduced later for the W$_x$B$_y$ configurations, $\alpha$-trigonal and $\beta$-trigonal, taking five random frames during the creation of the first deposited layer. Thicker boron-layer depositions were included at the end on the four relevant tungsten surfaces to data for the natural formation of borophene-like boron layers \cite{borophene-2D,FengBoro} and boron reconstructions \cite{bororec66,boroRec}. The last depositions had a boron layer thickness of around 10 Å.

To sample many-body repulsion and local atomic environments that are representative of sputtering simulations, we created surface structures were near-surface atoms are increasingly displaced. Specific atoms from the top first and second surface layers were displaced upwards to simulate an atom leaving the surface during sputtering. The first training contained frames up to 2 Å from the original top layer and spaced by 0.5 Å, and up to 4 Å from the second layer spaced by 0.75 Å, for each relevant phase. For the second training iteration, additional frames for $\alpha$-trigonal and $\beta$-trigonal were added in places where we found that the potential deviates most from DFT (although the second iteration did not produce much improvement as discussed in Validation section).

\section{Validation}

\subsection{Dimers}

\begin{figure*}[t] 
\centering
\includegraphics[width=0.9\linewidth]{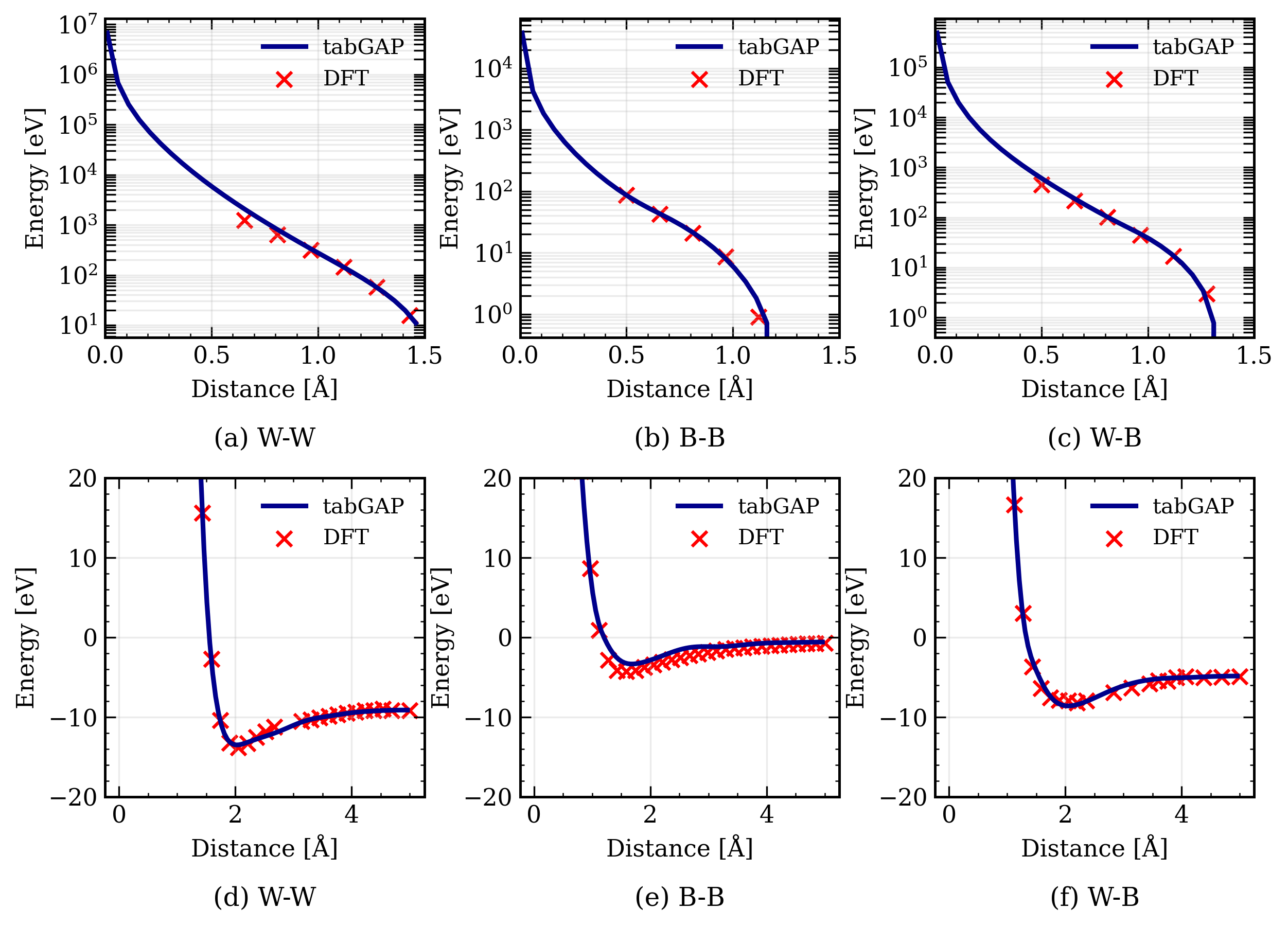}
\caption{Dimer curves of tabGAP and DFT with logarithmic energy scale for the first row representing short to very short distance (a-c) and the entire range in linear scale in the second row (d-f).}
\label{dimer-paper}
\end{figure*}

The dimer curves from tabGAP and DFT in Fig.~\ref{dimer-paper} show both the short distance repulsive part in the first row in logarithmic scale for the $y$-axis and the entire bond length range with equilibrium and attractive regions in the second row. All three dimers show overall excellent agreement with DFT across many orders of magnitudes, although the B-B equilibrium bond strength is somewhat underestimated. The repulsion down to very short distances is clearly well described by the potential, which makes it suitable for high-energy radiation damage simulations.

\subsection{Bulk properties}

\begin{figure*}[t] 
\centering
\includegraphics[width=0.9\linewidth]{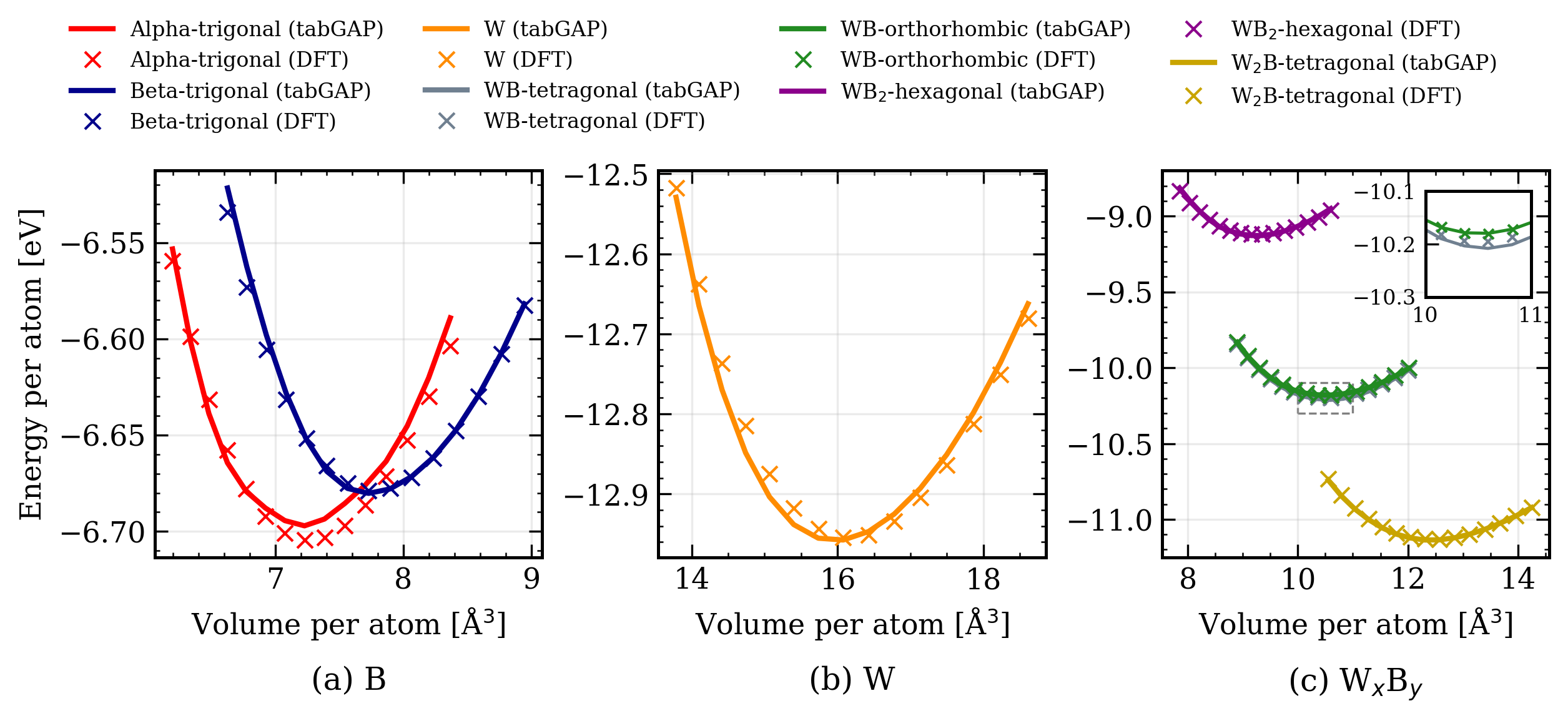}
\caption{Energy-volume curves for tabGAP and DFT for each relevant phase of boron (a), tungsten (b), and tungsten-borides (c).}
\label{EV-bulk}
\end{figure*}

The energy-volume curves and bulk moduli of the relevant phases were calculated and compared between the tabGAP and DFT. The bulk modulus was calculated by performing single point DFT or tabGAP calculations on isotropically scaled unit cells with a factor of $\pm$ 5\% from the equilibrium lattice spacings. Energies and volumes were extracted and fitted with an equation of state of Birch-Murnaghan type \cite{Birch}. The energy-volume curves in Fig.~\ref{EV-bulk} generated by tabGAP are overall in good agreement with DFT and correctly capture the phase ordering and elastic response. For boron, $\alpha$ is more stable than $\beta$ at 0 K and the smaller equilibrium volume shows that $\alpha$-B has a denser bonding compared to $\beta$-B. $\beta$-B will become more stable at higher temperatures or under very low pressures \cite{diagramphaseB}. The bulk moduli presented in Table~\ref{tab:bulkmod} show good agreement with DFT, although with small overestimation of around 5\%, with the exception of B$_2$W, where tabGAP overestimates the DFT value by about 10\%.

\begin{table}[h!]
\centering
\caption{Bulk moduli from tabGAP compared to DFT for different phases.}
\label{tab:bulkmod}
\begin{tabular}{lccc}
\hline
Material & $B$ [GPa] & $B_{\mathrm{DFT}}$ [GPa]\\
\hline
$\alpha$-B & 243 & 237 \\
$\beta$-B  & 230 & 215 \\
W          & 321 & 304 \\
BW         & 334 & 343 \\
BW$_2$     & 332& 330 \\
B$_2$W     & 365 & 324 \\
\hline
\end{tabular}
\end{table}

In Table~\ref{tab:latcst} the relaxed lattice constants from tabGAP and DFT obtained from energy minimizations, allowing for anisotropic cell changes, are presented. Overall, the lattice constants are in good agreement with DFT with $\leq$ 1.5\% errors for pure elements. However, BW shows a larger difference with an overestimation of around 4\% in the $y$-dimension.

\begin{table}[h!]
\centering
\caption{Lattice constants of different phases from tabGAP compared with DFT.}
\label{tab:latcst}
\begin{tabular}{lccc}
\hline
Material & $L_{\mathrm{x}}$ / $L_{\mathrm{y}}$ / $L_{\mathrm{z}}$ [Å] & $L_{\mathrm{DFT}}$ [Å] \\
\hline
$\alpha$-B & 4.84/4.84/12.62 & 4.9/4.9/12.55 \\
$\beta$-B  & 10.90/10.90/23.54 & 10.92/10.92/23.73 \\
W          & 3.17/3.17/3.17 & 3.19/3.19/3.19 \\
BW         & 3.07/8.83/3.14 & 3.18/8.51/3.10  \\
BW$_2$     & 5.59/5.59/4.76 & 5.58/5.58/4.80 \\
B$_2$W     & 3.02/3.02/14.10 & 3.02/3.02/14.06 \\
\hline
\end{tabular}
\end{table}

\begin{figure}[t] 
\centering
\includegraphics[width=0.7\columnwidth]{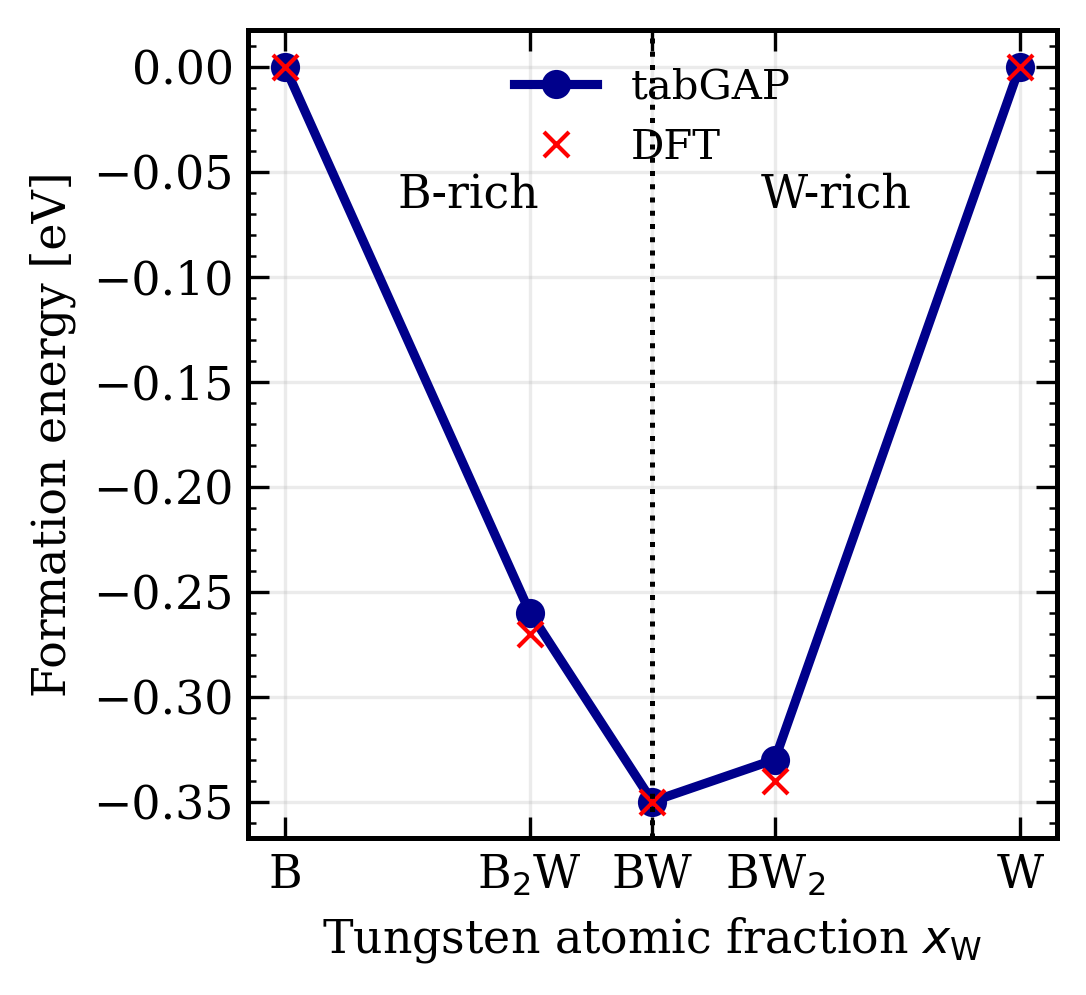}
\caption{Bulk formation energies per atom of tungsten borides for tabGAP compared to DFT.}
\label{bulkformE}
\end{figure}

The cohesive energies for the relevant elemental phases are given in Table~\ref{tab:cohesE} for tabGAP and DFT. Both DFT and tabGAP cohesive energies are calculated with their respective isolated atom energies (which are non-zero). To validate that the tabGAP predicts the correct thermodynamic stability of mixed W-B phases, we compute and show the bulk formation energies for tungsten borides in Fig.~\ref{bulkformE}. The tabGAP agrees well with DFT for cohesive and formation energies, with differences of less than 0.02 eV. The negative values for the bulk formation energies show that the compounds are favourable to form and thermodynamically stable, with lowest formation energy for the equiatomic BW phase in both tabGAP and DFT. Fig.~\ref{bulkformE} also shows that tungsten borides tend to favour W-rich or equiatomic compositions over B-rich ones in terms of thermodynamic stability as suggested in Ref.~\citenum{W2B}. In a W-rich environment, metallic bonds are dominant and the structure is closer to BCC, which leads to more negative potential energy per atom but higher volume.

\begin{table}[h!]
\centering
\caption{Cohesive energy per atom for pure elements. The isolated atom energies are B: $-0.296$ eV, W: $-4.565$ eV for tabGAP and B: $-0.287$ eV, W: $-4.571$ eV for DFT.}
\label{tab:cohesE}
\begin{tabular}{lccc}
\hline
Material & $E_{\mathrm{coh}}$ [eV] & $E_{\mathrm{coh}}^{\mathrm{DFT}}$ [eV] \\
\hline
$\alpha$-B & --6.40 & --6.42 \\
$\beta$-B  & --6.39 & --6.39 \\
W          & --8.39 & --8.38 \\
\hline
\end{tabular}
\end{table}

\subsection{Surfaces}

Boron adsorption energies on pure tungsten surfaces are presented in Table~\ref{tab:adsE}. The relevant sites for \hkl(100) are 4F, placed at the center of a square formed by four tungsten atoms, SB, which represents a bridge between two closer tungsten atoms, and LB that represents a bridge from further atoms. The relevant adsorption sites for \hkl(110) are also LB, SB, and TOP, where the last represents the site directly on top of a surface atom. The \hkl(111) surface has different sites such as FCC representing a 3F site that forms an FCC-like stacking underneath, BL$_{32}$ which represents a bridge-like low-coordination site between surface atoms, and a TOP site. The \hkl(112) surface has 3F, SB, and TOP as shown in Ref. \citenum{adsW} for W adatoms, also showing the 3F position as the most stable. The DFT reference values are taken from previous work for \hkl(100), \hkl(110), \hkl(111) \cite{adsorp}. Static relaxations using the tabGAP potential were performed to find and compute the adsorption site energies, consistent with the DFT reference \cite{adsorp}. 

The adsorption energies for tabGAP shown in Table~\ref{tab:adsE} agree reasonably well with DFT. Larger values mean more stable sites. While there is a systematic underestimation, most importantly the order of stability is consistent with DFT. The fact that all the values are underestimated and not scattered suggest that the potential is consistent. For some cases, such as the bridge sites for \hkl(100) and the TOP site for \hkl(112), a transition during minimization occurs to a much more stable site.

\begin{table}[h]
\centering
\caption{Adsorption energies of boron on each relevant pure tungsten surface compared between tabGAP and DFT reference values from Ref. \cite{adsorp}.}
\label{tab:adsE}
\begin{tabular}{c c c c}
\hline
Surface & Site & $E_{\mathrm{ads}}$ [eV] & $E_{\mathrm{ads}}^{\mathrm{DFT}}$ [eV] \\
\hline
\hkl(100) & 4F & 7.57 & 7.80  \\
\hkl(100) & LB & LB $\rightarrow$ 4F & 5.72 \\
\hkl(100) & SB & SB $\rightarrow$ 4F & 5.34 \\
\hline
\hkl(110) & LB & 5.65 & 6.58 \\
\hkl(110) & SB & 4.31 & 5.46 \\
\hkl(110) & TOP & 2.81 & 3.72 \\
\hline
\hkl(111) & BL32 & 6.15 & 6.45 \\
\hkl(111) & FCC & 6.15 & 6.42 \\
\hkl(111) & TOP & 2.50 & 2.92 \\
\hline
\hkl(112) & 3F  & 6.17 & -- \\
\hkl(112) & SB  & 3.80 & -- \\
\hkl(112) & TOP & TOP $\rightarrow$ SB & -- \\
\hline
\end{tabular}
\end{table}

Surface energies presented in Table~\ref{tab:surface_energy} were calculated for all relevant phases. For the tabGAP, each bulk phase was minimized first by using anisotropic expansion to zero pressure, after which a vacuum layer (30 Å) was added and the atom positions were minimized again but with no box size change. The difference in energy between the bulk and the surface sample divided by two times the area gives the surface energy of each configuration. For DFT, the surface energies were calculated in the same way. For simplicity, we use the cubic notation to denote the surfaces of all sides of the conventional unit cells for each structure. Table~\ref{tab:surface_energy} shows that the tabGAP agrees well with DFT and reproduces the correct order of stability for all phases. There is a consistent underestimation of 0.01--0.06 eV/Å$^2$, similarly to the surface adsorption energy.

\begin{table}[ht]
\centering
\caption{Surface energies of boron, tungsten, and B-W compounds.}
\label{tab:surface_energy}
\begin{tabular}{l c c l}
\hline
Material & $\gamma$ [eV/Å$^2$] & $\gamma_{\mathrm{DFT}}$ [eV/Å$^2$] \\
\hline
$\alpha$-B \hkl(001) & 0.28 & 0.32 \\
$\alpha$-B \hkl(100) & 0.24 & 0.27 \\
$\alpha$-B \hkl(010) & 0.24 & 0.27 \\
\hline
$\beta$-B \hkl(001) & 0.27 & 0.30\\
$\beta$-B \hkl(100) & 0.22 & 0.25 \\
$\beta$-B \hkl(010) & 0.22 & 0.25 \\
\hline
BW \hkl(001) & 0.22 & 0.23  \\
BW \hkl(100) & 0.23 & 0.24 \\
BW \hkl(010) & 0.19 & 0.18  \\
\hline
W \hkl(100) & 0.25 & 0.27 \\
W \hkl(110) & 0.20 & 0.19 \\
W \hkl(111) & 0.23 & 0.25 \\
\hline
B$_2$W \hkl(001) & 0.34 & 0.4 \\
B$_2$W \hkl(100) & 0.18 & 0.21 \\
B$_2$W \hkl(010) & 0.18 & 0.21 \\
\hline
BW$_2$ \hkl(001) & 0.19 & 0.22 \\
BW$_2$ \hkl(100) & 0.23 & 0.24 \\
BW$_2$ \hkl(010) & 0.23 & 0.24 \\
\hline
\end{tabular}
\end{table}

\subsection{Repulsive potential and decohesion}

To test the repulsive part of the potential and produce local atomic environments similar to ones expected in sputtering simulations, we performed quasi-static drag calculations of surface atoms in both DFT and tabGAP. In these quasi-static drag calculations, a single atom is moved incrementally toward the vacuum, emulating an atom being sputtered. Both DFT and tabGAP were used for all relevant phases with orientation \hkl(001) for B and WB, and \hkl(100) for W. Two separate cases were done where an atom from the topmost layer and one from the first subsurface layer was dragged upwards. The subsurface atom drag probes repulsion as the atom passes nearby the surface-layer atoms. For BW, the calculations were done once for a tungsten atom and once for a boron atom on the top surface layer. For tabGAP, the range was chosen from the original atom position to the surface layer position plus 5 Å, with 100 points. For DFT, the range was the same, but only 20 points were sampled. The potential energy evolution of the system was plotted as a function of the atom displacement for direct comparison between tabGAP and DFT.

\begin{figure*}[t] 
\centering
\includegraphics[width=0.99\linewidth]{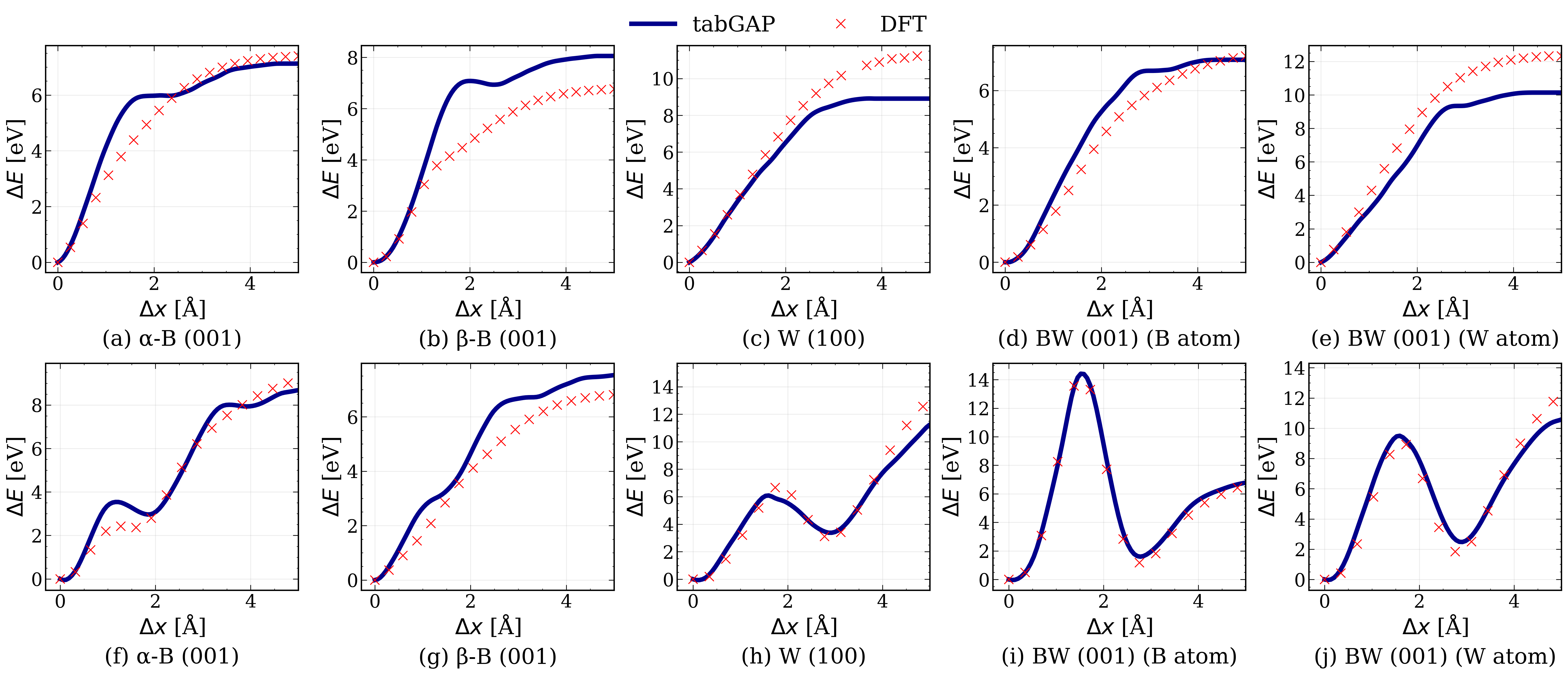}
\caption{Quasi-static drag of top-layer (a-e) and second-layer (f-j) atoms upwards along the surface normal.}
\label{quasi-static}
\end{figure*}

The results are shown in Fig.~\ref{quasi-static} and while some cases show excellent agreement between tabGAP and DFT, there are also notable disagreements both in the energy profile and final decohesion limits. For example, for pure B there are local minima in the tabGAP that do not appear in DFT. Attempts to improve the potential by adding the structures to the training database did not help to circumvent the problem. The drag curves for pure W and BW do not show similar discrepancies and are overall in good agreement with DFT. Important to note is that the tabGAP is trained with correct spin-polarised isolated atoms, which is critical for reproducing experimental cohesive energies better, while the DFT drag curves in Fig.~\ref{quasi-static} are from non-spin-polarised DFT calculations. This means that there is an expected mismatch between tabGAP and DFT in the decohesion limits. The mismatch for W atoms is around 3 eV, which explains the decohesion differences for the W-atom drag curves and should not be considered a failure of the trained tabGAP. For B atoms, however, the effect of spin-polarisation is small and does not account for the tabGAP discrepancies.

\subsection{Melting and liquid properties}

Melting temperatures for the relevant phases were simulated using the liquid-solid coexistence interface method \cite{meltcohex} with an increment of 100 K. All melting temperatures agree well with experiments, underestimating it by around 100-150 K or $\leq 5$\%. $\alpha$-B is known to turn into $\beta$-B due to a phase transition at around 1000 K \cite{alphatrans}, and melting simulations starting from both phases showed the same melting point, 2250 $\pm~50$ K.

\begin{table}[h!]
\centering
\caption{Melting temperatures from the liquid–solid coexistent method compared with experimental values.}
\label{tab:tmelt}
\begin{tabular}{lcc}
\hline
Material & $T_\text{melt}$ [K] & $T_\text{melt}^\text{exp.}$ [K] \\
\hline
$\beta$-B  & 2250 $\pm~50$ & 2349 \cite{meltWandB}  \\
B--W       & 2850 $\pm~50$ & 2938 \cite{meltWB} \\
W          & 3525 $\pm~50$ & 3695 \cite{meltWandB} \\
\hline
\end{tabular}
\end{table}

\section{Applications}

\subsection{Method}
\label{application}

LAMMPS was used for all simulations, with an adaptive time step implemented \cite{KAItimestep}, with a maximum of 2 fs when only tungsten is involved and 0.5 fs when boron is involved as either a material or an incoming ion. The boundaries were periodic in the $x-$ and $y-$directions to represent an infinite surface and open in the $z-$direction. Three different layers were defined within the simulation box. The first layer at the bottom consisted of fixed atoms to prevent the lattice from drifting and the atoms from leaving the cell from the bottom because of open boundaries in the $z-$direction. The second layer consisted of thermalised atoms controlled by a Nose-Hoover thermostat at 300 K to emulate heat losses to the bulk, and all other atoms above the previous defined layers followed the NVE ensemble. Although 300 K is under the typical range of fusion reactor operation, many experiments are done at room temperature. Additionally, as ballistic effects are the dominating factors to influence the sputtering and temperature is usually a minor contributor to sputtering overall, the chosen temperature will not significantly affect the trends. The interactions between B and W are described by our current tabGAP potential, which include a short-range repulsive corrections to accurately capture high-energy collision event. The electronic stopping power was neglected due to the quite low impact energies considered ($\leq$ 1 keV).

First, a relaxation simulation of 6 ps at 300 K was performed on all of these surfaces. The impact events following the relaxation lasted 5 ps for all simulations. For deposition, a 6 ps relaxation was done after each impact simulation to ensure that there was no temperature build-up during the series. The fixed atoms and the thermalised layer were about a few layers thick. For statistics, 250 single impacts on 6 different thermalised cells totalling 1500 single impacts were done for each parameter set. W and B were chosen as incoming ions with three energies (200 eV, 500 eV, and 1000 eV) that were selected to cover relevant sputtering regime from just above the sputtering threshold to the upper part of the low-energy linear cascade regime. This range captures the transition from threshold-sensitive sputtering at low energies to more cascade driven behavior at higher energies. These ions were used to sputter W(\hkl(100), \hkl(110), \hkl(111) and \hkl(112)), WB, $\alpha$- and $\beta$-B, and their initial point was set at 7 Å to ensure that they were outside the cut-off zone of the potential. The size of the box was energy-dependent and carefully chosen to balance computational efficiency without compromising the results. The size of the box in the $z-$direction was larger than in the lateral $x-$ and $y-$directions to minimize and prevent ions from channelling through the material and leave the open boundary at the bottom. The box sizes for 200 eV, 500 eV, and 1000 eV single impact on tungsten in the $x-$, $y-$directions were around $5 \times 5 $ nm$^2$, $5 \times 5 $ nm$^2$ and $6 \times 6 $ nm$^2$. For $\alpha$ were around $3 \times 2.5 $ nm$^2$, $3 \times 2.5 $ nm$^2$ and $4 \times 3.5 $ nm$^2$, for $\beta$ around $3.3 \times 2.8 $ nm$^2$, $3.3 \times 2.8 $ nm$^2$ and $4.2 \times 3.8 $ nm$^2$, for BW around $3.8 \times 3.3 $ nm$^2$, $3.8 \times 3.3 $ nm$^2$ and $5.6 \times 5.1 $ nm$^2$. The $z-$direction was set differently depending of the incoming ion, the energy, and the material to keep the channelling effect under a threshold of 5\%.

The deposition is performed the same way as in the sputtering simulation with the exception of the energy, angle of incoming ion, and number of atoms. The cells were 8 unit cells in all directions for \hkl(100) and equivalent sizes for \hkl(110), \hkl(111), and \hkl(112), making it around 1000 atoms. The boron is deposited with an energy of 5 eV to stick to the surface and with a small incoming angle of 10 degrees to avoid potential channelling directions. A total of 2000 depositions were conducted for each case and an adaptative surface detection based on the highest atom in the $z-$position was used to always respect the ion initialization point of 7 Å above the surface.

\subsection{Results \& Discussion}

\begin{figure*}[t] 
\centering
\includegraphics[width=0.9\linewidth]{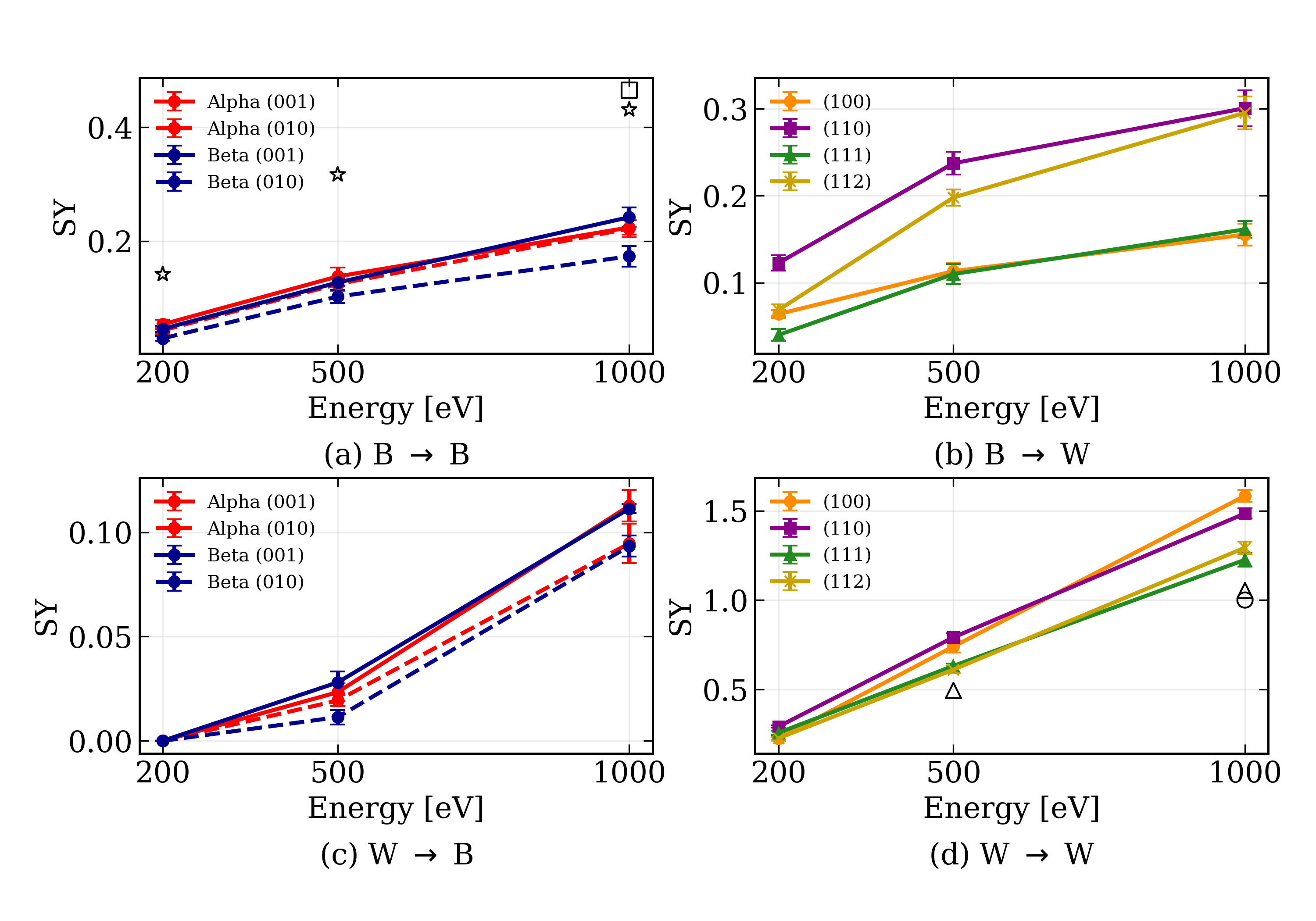}
\caption{Sputtering yield of boron (first column) and tungsten (second column) by boron (first row) and tungsten (second row), compared with experimental data from Refs. \citenum{HECHTL90} (square), \citenum{HECHTL92} (circle), \citenum{ECKSTEIN93} (triangle), and \citenum{YAMAMURA95} (star).}
\label{SI-SY}
\end{figure*}

For both B and W ions, the sputtering yields of both boron and tungsten surfaces, shown in Fig.~\ref{SI-SY}, are close to zero at 200 eV indicating that we are near to the threshold energy regime. At 500 eV, sputtering become more prominent across all system marking the onset of developed cascade formation. At 1000 eV, all combinations have a huge increase compare to 200 eV with W-W showing the largest increase in sputtering yield. The complete data about sputtering yield, reflection yield, and channelling are available in the supplementary material Tables II - IV. In the sub-keV regime, the sputtering yield is increasing with the increase of energy for all combination which is consistent with the growth of collision cascade. This increase is due to a deeper penetration under the surface and an enhanced momentum transfer. Based on sputtering theory \cite{SigmundTheory}, after a certain energy, each materials will have cascades too deep and will cause a decrease in sputtering yield since insufficient amount of energy will reach the surface and contribute to sputtering. This regime where sputtering yield decrease with the increase of energy is usually far from the sub-keV regime. However, we can see that for the lighter element B, the yield is not increasing that much more when increasing the energy from 500 eV to 1000 eV, whereas for W the increase is more substantial. This suggests that the heavier element W will deposit the energy in the near surface layers causing more sputtering. Our study show three regimes of the sputtering, the threshold regime at around 100 eV -- 200 eV where the energy is just high enough for sputtering, the transition regime where small cascades start to develop, and the cascade-dominant regime at around 1000 eV where the collision cascades are fully developed and the recoils reach the surface. Additionally, for B as the ion also the channeling regime starts to emerge, where the ion penetrates deep into the material and mainly causes bulk damage.


Regarding all four pairs of ion-target, self-sputtering (W-W, B-B) consistently exhibit a higher sputtering yield than the cross-species (B-W, W-B) on the same target. W-W is showing the highest yield due to matched masses giving efficient momentum transfer and low to no reflection, thus producing denser and more developed cascades. The yield of B impacting W is lower than W on W because of the masses mismatch and the higher reflecting yield, as well as a deeper possible penetration. The same pattern is visible for B target with B-B higher than W-B due to the same less efficient energy transfer for cross-species. B-B and B-W remain comparable with the same order of magnitude.

Surface orientation play a significant role for tungsten sputtering, especially when boron is used as the incoming ion. For boron ions the close-packed surface \hkl(110) shows the highest sputtering yield at all energies and the more open directions show a significantly lower sputtering yield. This can be understood that for the close packed surface the B ions are able to transfect energy to the surface layers of the material, therefore causing sputtering. For the more open surface orientations, the boron ions, due to their low mass compared to W, can penetrate deep into the material, causing bulk damage instead of sputtering. For W ions, due to their more efficient energy transfer to the surface, the difference is not as visible. The same mechanisms are present at higher energies, but at the lowest energy, all surface orientations sputter in a similar manner. This is due to that at this regime all energy from the incoming W ion is deposited in the surface layers of W, and therefore the sputtering yield is similar. Unlike tungsten, surface orientation of boron  has a lower impact on the sputtering yields. For both tungsten and boron as the incoming ion, the difference between \hkl(001) and \hkl(010) are small. This indicates that crystallographic effects are less significant in low mass, covalent systems where cascades are less dense and less efficiently propagated than for metals. 

For boron self-sputtering, the sputtering yields do not agree with previous reported experiments with an underestimation of a factor of 2 lower than experiments. However, the trend is correctly reproduced for this energy range. For tungsten self-sputtering, the simulation are closer to the experimental sputtering yields, but are overestimated by around 50\%, however again showing the correct trend. The difference seen between our potential and the sputtering experiment might be due to different factors, such as the sample surface conditions and/or lack of electronic stopping power in our simulation. The electronic stopping could affect the sputtering to some extent, where the losses can be different in light and heavy materials due to different ion masses. Another factor not accounted for here is surface roughness, which is always present experimentally. It is known that for W surfaces it will lower the net erosion, due to prompt redeposition of the sputtered particles. For the lighter and less densly packed boron, possible surface roughness and/or surface amorphisation could affect the experimental sputtering yields. The only MD study on boron sputtering are from a Neural network potential for BO$_x$ \cite{NNP25} showing good agreement with boron self-sputtering experiments at 200 eV, and a Reactive force field (ReaxFF) \cite{reacB} showing sputtering yield at lower energy than those studied. On the other hand, self-sputtering of tungsten using MD simulations have been widely studies. Some Tersoff potential show good agreement with the simulation at 200 eV \cite{faithWW}. There is no study on sputtering of tungsten-boron, tungsten on boron, or boron on tungsten due to the lack of potentials for these configurations. Our potential will allow to investigate sputtering of any tungsten and boron combination by both B and W ions.

\begin{figure*}[t] 
\centering
\includegraphics[width=0.99\linewidth]{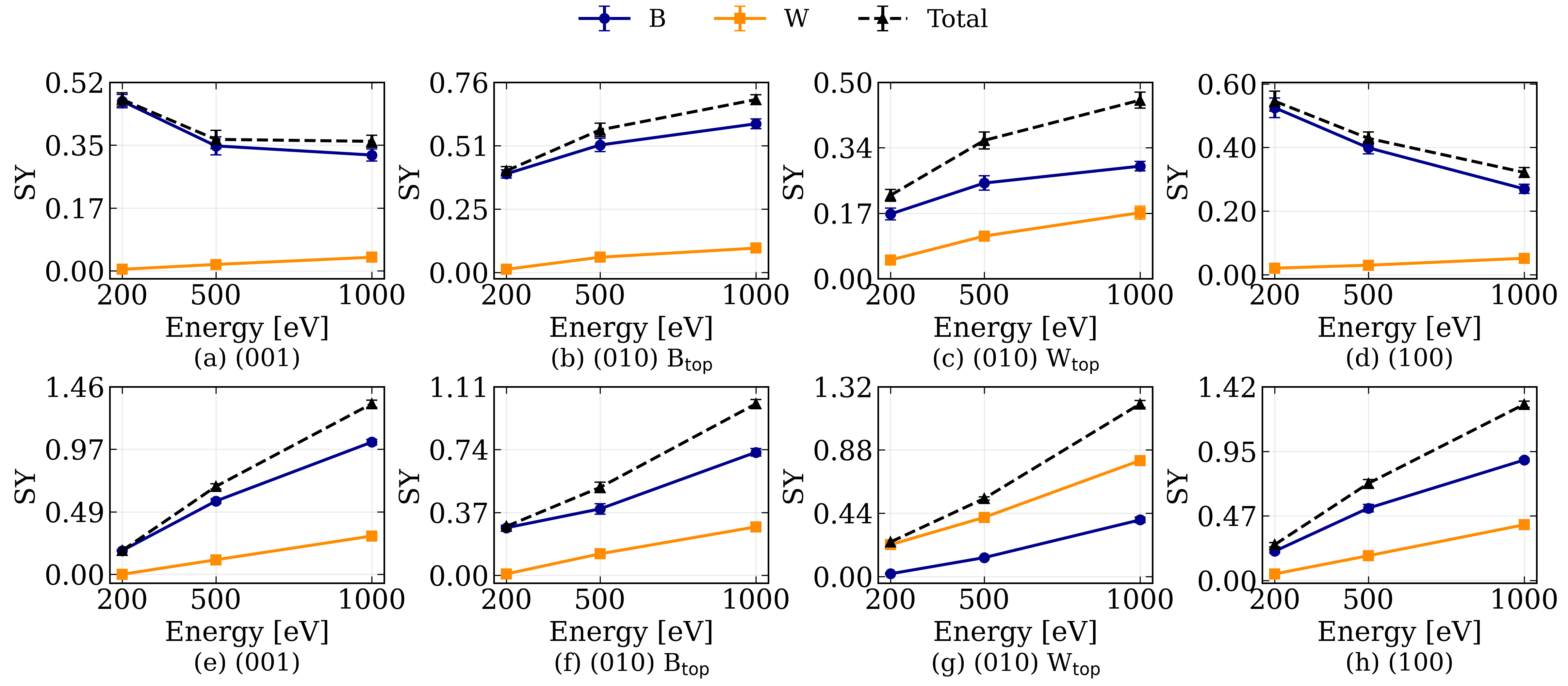}
\caption{Sputtering yield and elemental sputtering yield of WB by B ions in the first row and W ions in the second row.}
\label{SI-BW-SY}
\end{figure*}

Looking at boron sputtering of mixed WB surfaces (Fig.~\ref{SI-BW-SY}), we observe both differences and similarities in the trends in relation to boron impacting elemental materials (Fig.~\ref{SI-SY}). We see that in Fig.~\ref{SI-BW-SY}(a-d) the trends are different with some reduction in sputtering yield at higher energy that is absent in the pure elemental targets. The observed channelling rate through the box was very low, under 3.5\%. The complete data are available in the supplementary material Tables V - VII, containing sputtering yields, reflection yields, and channelling rates. The light mass of the boron ion limits momentum transfer to the heavier tungsten sublattice, keeping the sputtering yield of tungsten species lower than that of boron. This is supported by the high reflection yields of boron ions impacting WB surfaces, suggesting that many ions will be reflected rather than contributing to energy transfer to the lattice and subsequent sputtering. The surface \hkl(010)W$_{\mathrm{top}}$ shows the highest reflection yield, which is consistent with ions of low mass that are reflected by surfaces consisting of high mass elements. The boron sputtering yield is reduced at higher energy for \hkl(001) and \hkl(100) surfaces because the boron ions tend to travel deeper in the bulk making it inefficient to sputter atoms by depositing their energy far below the surface. For tungsten impacting the WB surfaces, Fig.~\ref{SI-BW-SY}(e-h), the trends are similar to elemental materials, Fig.~\ref{SI-SY}. The tungsten ions have a near zero reflection yield, meaning they either penetrate deep into the material or stick to the surface. In both cases, all the energy of the ion is deposited in the material. Boron is preferentially sputtered in all cases with the exception of the \hkl(010)W$_{\mathrm{top}}$ surface, where the topmost layer is consisting of only tungsten atom. This preferential sputtering will for both ions mean that during prolonged irradiation the surface will be tungsten enriched at the top.

\begin{figure*}[t] 
\centering
\includegraphics[width=0.8\linewidth]{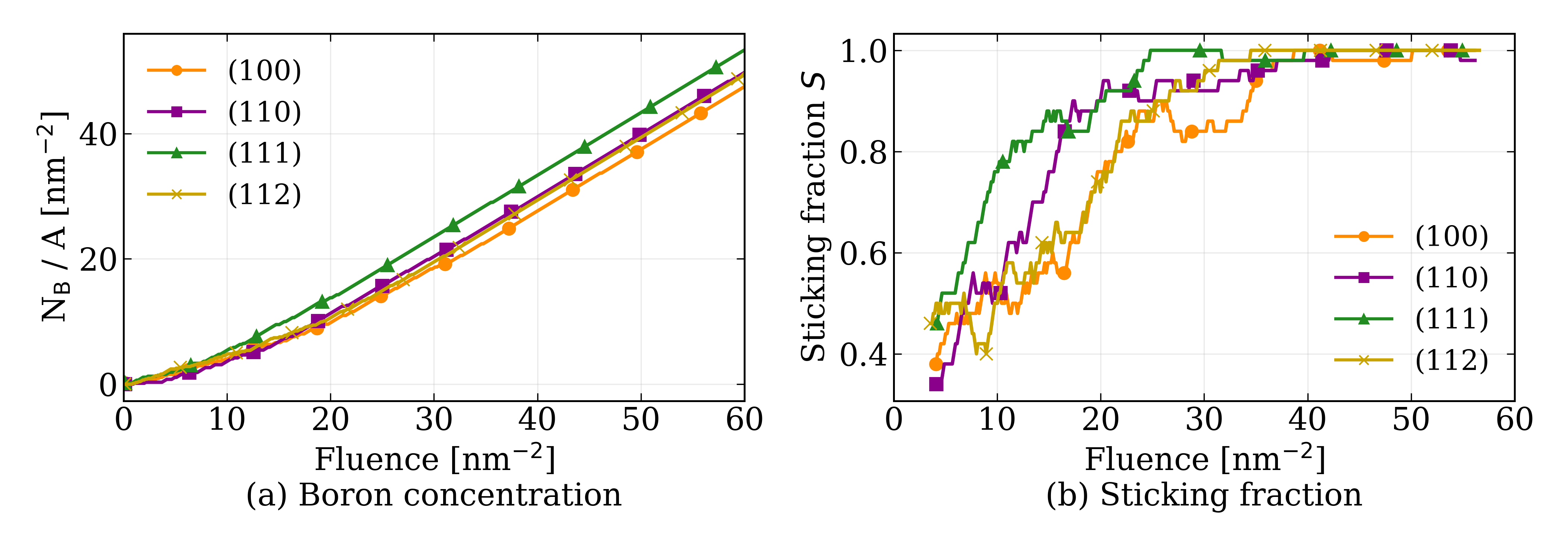}
\caption{Boron concentration on top of different tungsten surfaces during deposition (a), and the corresponding sticking fraction (b).}
\label{depo_rate}
\end{figure*}

Fig.~\ref{depo_rate} shows the accumulation of boron per surface area and the corresponding sticking probability during deposition. It can be seen that, at the right beginning of the deposition, boron atoms stick less when deposited over the tungsten surface. As the deposition fills the surface with boron, the surface becomes richer in boron. The deposition rate becomes stable when the surface is fully boronized and most of the boron sticks after deposition. The \hkl(111) surface has a higher sticking probability during the construction of the first layer than other surfaces, with \hkl(100) and \hkl(112) the lowest. After a few tens of ions per nm$^2$, the sticking fraction is stabilized, where all surfaces show a fraction close to 1. After the first boron layers have been deposited, the surface orientation of the tungsten does not matter any more. The difference in the sticking fraction (Fig.~\ref{depo_rate}b) can also be seen in Fig.~\ref{depo_rate}a, there the \hkl(111) surface show a higher number of B atoms per area, due to the higher sticking fraction initially.

\begin{figure}[t] 
\centering
\includegraphics[width=0.9\columnwidth]{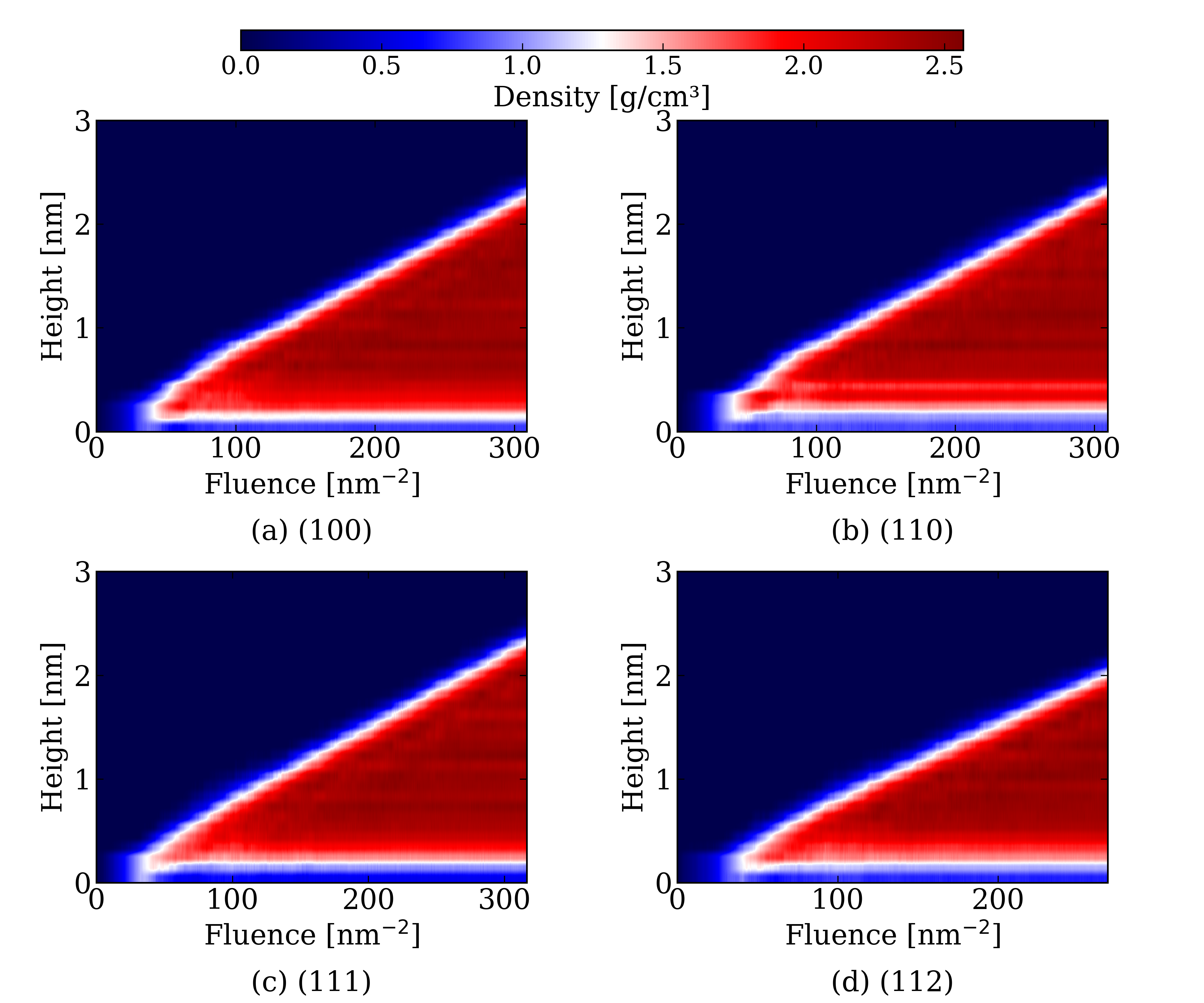}
\caption{Heat map showing the density of deposited boron on top of different tungsten surfaces.}
\label{density_cut}
\end{figure}

We computed the density of two different boron phases. We found a density of 2.48 g/cm$^3$ for $\alpha$ boron and 2.32 g/cm$^3$ for $\beta$ boron, with a experimental value reference of 2.46 g/cm$^3$ \cite{densityAlphaB} and 2.35 g/cm$^3$ \cite{densityBetaB}, showing good agreement between our tabGAP and experimental values. The density profiles in Fig.~\ref{density_cut} have been computed over the entire deposition of tungsten \hkl(100), \hkl(110), \hkl(111), and \hkl(112) surfaces. To compute the density of the deposited layer, a slab has been introduced with a value of $\pm$2 to the height value. From this 4 Å thick slab, the volume, number of boron atoms, and the mass of boron have been used to calculate the local density. The 0 height in the plots corresponds to the lowest position of the boron atoms. Only the density for boron atoms was computed. Approximately at a fluence of 50 $nm^{-2}$, when most of the relevant adsorption sites are filled, a layer similar to that of borophene is elevated above the surface. This layer is consistent with experiments showing that the boron prefers to form a 2D structure first rather than bulk \cite{borophene-2D,FengBoro}. This layer is more prominent for the surface \hkl(110), shown in the supplementary material Fig. SI. After sufficient  deposition, this borophene-like layer start to be closer to bulk-like material giving an amorphous boron with a density close to that of the relevant phases. In supplementary material Fig. SI, we can see a few isolated atoms occupying the site inside the tungsten layer, then the first layer above the tungsten surface, and the second layer elevated with some atom closing the gap and bridging the two layers. However, it is very difficult to determine whether these layers are real borophene phases or just planar voids inside the newly deposited layer. Boron is known to be vacancy rich with space between B12 icosahedra clusters \cite{Galeev-boro}. After more deposition, these layers slowly vanish, showing a density close to the original density $\alpha$ and $\beta$ \cite{transformation-bulk}. 

\begin{figure*}[t] 
\centering
\includegraphics[width=0.99\linewidth]{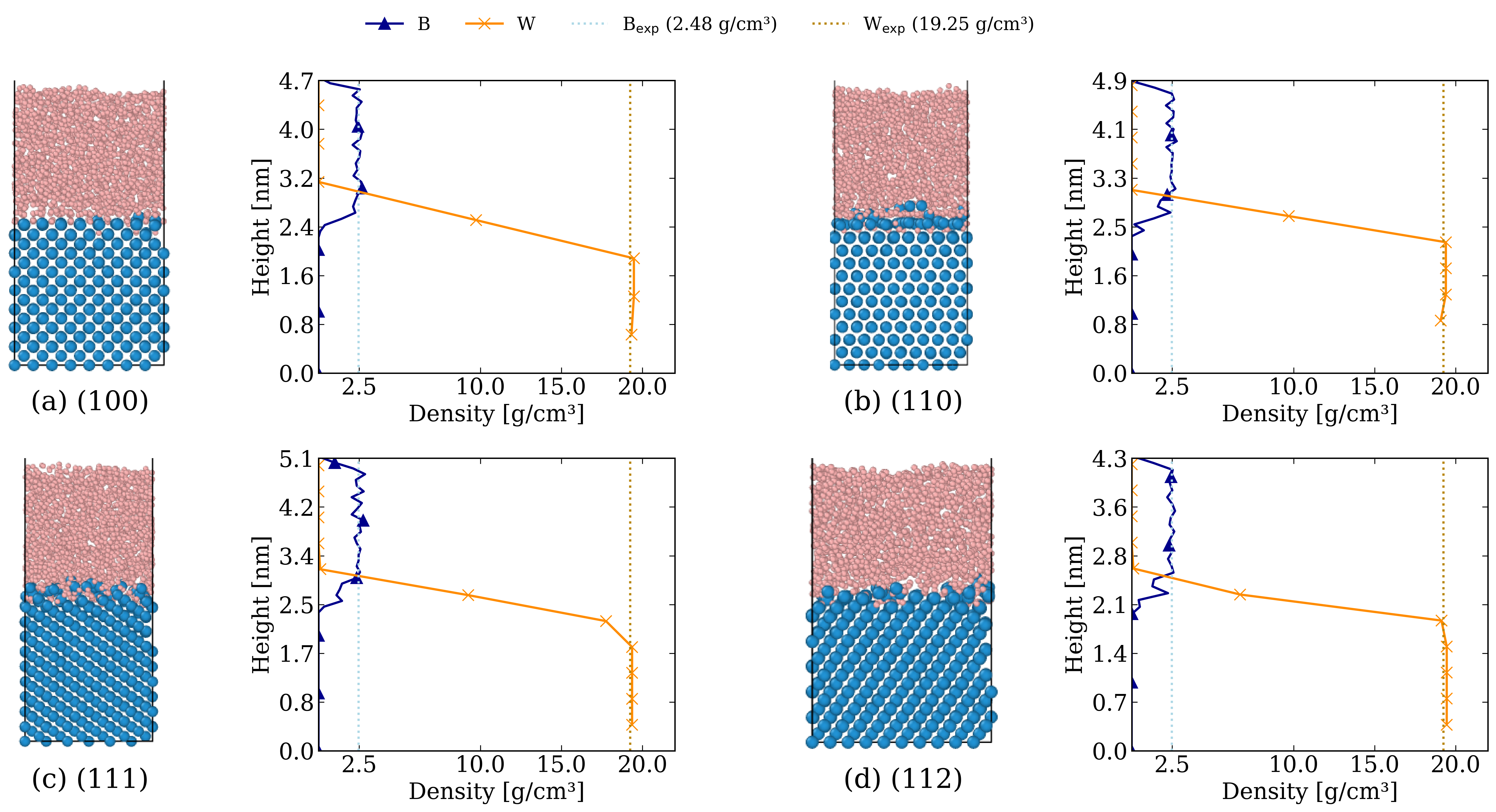}
\caption{Snapshots representing the structure after 2000 boron deposition events for (a) \hkl(100), (b) \hkl(110), (c) \hkl(111) and (d) \hkl(112) tungsten surfaces, with their respective density profiles on the right side. The experimental value for B is the $\alpha$ phase taken from Ref. \citenum{densityAlphaB} and for W taken from Ref. \citenum{densityW}.}
\label{combine-frame}
\end{figure*}

Fig.~\ref{combine-frame} shows the boron deposited on the surfaces after 2000 iterations. A different scanning thickness has been chosen for W (4-6 Å) and B (1 Å) because tungsten has a larger volume/atom than boron, thus introducing larger statistical fluctuations in the density profile. The data of the very bottom of the box has been removed due to border effects. The deposited boron is showing an amorphous dense phase with a density of around 2.42--2.44 g/cm$^3$. These values lie between the $\beta$ and $\alpha$ phases making it closer to the $\alpha$ phase. It has been reported that deposited boron form an amorphous phase with $\alpha$-rhombohedral short-range ordering \cite{amourphalpha}. To conclusively determine the structure of any of those phases is visually complicated, however the deposited boron have a similar density as the crystalline ones, shows it to be densely packed. The B/W interface region is clearly visible in all panels as a transition zone where both densities overlap. This overlapping zone is around 6-8 Å and containing both boron and tungsten atoms. This zone is known as ion beam mixing or interfacial mixing \cite{ionMIX}, where ion irradiation cause atomic rearrangement inside the material which lead to mixing and alloying of the interface. Ion mixing seems to be less pronounced for the \hkl(100) W surface deposition compared to the other surfaces. The height for the same number of deposited atoms varies due to the sticking factor as seen previously in Fig.~\ref{depo_rate}b, where \hkl(110) and \hkl(111) is higher than \hkl(100) and \hkl(112). 

\section{Conclusions}

In order to study fusion relevant materials during irradiation and boronisation, an interatomic potential for tungsten and boron is required. We trained a general purpose machine learning potential based on a DFT training dataset, allowing high speed and accuracy in simulating large-scale systems using MD, within the tabGAP formalism. The potential showed very good agreement with DFT and most errors on the order of a few percent. Sputtering of crystalline pure B and W showed an increase with energy, where tungsten showed a pronounced surface orientation dependent sputtering and boron only a more subtle one. The trends are in good agreement with experiments, however, the magnitude is not well reproduced, most likely due to the ideal configurations used in the MD simulations. For WB alloys, boron tends to sputter more than tungsten with the exception of \hkl(010) with the W atom on top. Tungsten ions exhibit near-zero reflection, indicating efficient penetration and full cascade development, whereas boron ions are largely reflected. Under boron ion bombardment, the sputtering yield saturates or decreases due to the lighter boron ion penetrating deeper reducing near surface energy deposition.

The boron deposition on top of tungsten surfaces exhibits a two stage growth mechanism for all surface orientations. At first, boron has a reduced sticking factor due to deposition on bare tungsten, until a full first layer of boron is produced. Then the sticking factor goes close or equal to 1 as the second stage. During the boron deposition, a borophene-like structure is generated above the tungsten surface until enough deposited atom make the structure bulk-like, also experimentally seen when boron is deposited at a low rate. After some layers of deposited boron, a dense amorphous phase with the density close to the $\alpha$ and $\beta$ phases is formed. These simulations provide understanding of how boron grows on tungsten surfaces during boronisation, and pave the way for future large-scale simulations combining both deposition phases during boronisation, and sputtering on those surfaces during irradiation phase of fusion reactors.

\section*{Acknowledgements}

This work has been carried out within the framework of the EUROfusion Consortium, funded by the European Union via the Euratom Research and Training Programme (Grant Agreement No 101052200 -- EUROfusion). Views and opinions expressed are however those of the author(s) only and do not necessarily reflect those of the European Union or the European Commission. Neither the European Union nor the European Commission can be held responsible for them. Computer time granted by the IT Center for Science -- CSC -- Finland is gratefully acknowledged. JB acknowledges funding from the Research council of Finland through the OCRAMLIP project, grant number 354234.

\bibliography{article}

\end{document}


\textbf{\Large Supplementary Information for ``Machine-learned interatomic potential for sputtering of tungsten-boron surfaces"}

\section{Dataset proportion}


Table~\ref{tab:dataset-tabWB} shows the different types of configuration for the WB and B parts with their respective number of configurations and atoms in proportion to the entire potential training data. A similar Table for the W part can be found in Table II in Ref. \cite{jesperW}.

\section{Sputtering and reflection yields}

Tables~\ref{tab:dataset-sput-200ev} -~\ref{tab:dataset-sput-1000ev} represents the sputtering yield for boron \hkl(001), \hkl(010), \hkl(100) and tungsten \hkl(100), \hkl(110), \hkl(111), \hkl(112), for both W and B as incoming ions. Tables~\ref{tab:dataset-sputWB-200ev} -~\ref{tab:dataset-sputWB-1000ev} represents the sputtering yield for WB \hkl(001), \hkl(010) with B as top atoms, \hkl(010) with W as top atoms, and \hkl(100) for both W and B as incoming ions. Channelling events in \% are also disclosed for each configuration.

\section{Deposition}

\begin{figure}[hb!] 
\centering
\includegraphics[width=0.5\columnwidth]{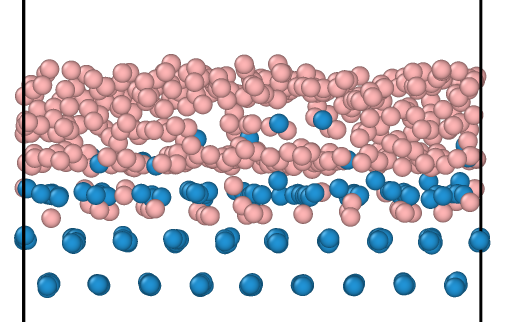}
\caption{Borophene-like structure elevation during deposition of B over tungsten \hkl(110)}
\label{boro-like-110}
\end{figure}

Fig.~\ref{boro-like-110} shows a borophene-like structure after deposition of B over \hkl(100) tungsten surface.

\bibliography{article}

\newpage
\clearpage

\begin{table}[htbp]
\centering
\caption{Configuration table for the B and WB parts}
\begin{tabular}{llrrrr}
\toprule
Group & Config type & Num. configs & Num. atoms & Proportion config & Proportion atoms \\
\midrule
Bulk & scaled\_structure\_B\_highE & 40 & 4590 & 0.8628127696 & 3.260244199 \\
  & scaled\_structure\_B\_lowE & 50 & 4770 & 1.078515962 & 3.388096912 \\
  & scaled\_structure\_B\_noEXP & 50 & 4470 & 1.078515962 & 3.175009056 \\
  & scaled\_structure\_WB\_highE & 50 & 1270 & 1.078515962 & 0.9020719243 \\
  & scaled\_structure\_WB\_lowE & 50 & 510 & 1.078515962 & 0.3622493554 \\
  & B\_thermal\_disp & 5 & 720 & 0.1078515962 & 0.5114108547 \\
  & WB\_thermal\_disp & 5 & 720 & 0.1078515962 & 0.5114108547 \\
  & MD-therm\_B & 20 & 4215 & 0.4314063848 & 2.993884379 \\
  & MD-therm\_WB & 20 & 2460 & 0.4314063848 & 1.74732042 \\
  & B\_dimer & 20 & 40 & 0.4314063848 & 0.02841171415 \\
  & WB\_dimer & 10 & 20 & 0.2157031924 & 0.01420585707 \\
\midrule
Surface & WB\_vacancies & 20 & 2740 & 0.4314063848 & 1.946202419 \\
  & MD-B\_adsorption & 6 & 838 & 0.1294219154 & 0.5952254114 \\
  & extd\_MD-B\_adsorption & 6 & 1074 & 0.1294219154 & 0.7628545249 \\
  & B\_deposit\_on\_W & 7 & 1186 & 0.1509922347 & 0.8424073245 \\
  & B-surface & 15 & 3415 & 0.3235547886 & 2.425650096 \\
  & W-vac & 4 & 538 & 0.08628127696 & 0.3821375553 \\
  & B2W\_WB-surface & 5 & 960 & 0.1078515962 & 0.6818811396 \\
  & BW2\_WB-surface & 5 & 720 & 0.1078515962 & 0.5114108547 \\
  & BW-ortho\_WB-surface & 5 & 1280 & 0.1078515962 & 0.9091748528 \\
  & BW-tetra\_WB-surface & 5 & 1280 & 0.1078515962 & 0.9091748528 \\
  & EXTD\_WB-surface & 8 & 1616 & 0.1725625539 & 1.147833252 \\
  & surfaceE & 15 & 2133 & 0.3235547886 & 1.515054657 \\
  & B\_deposit\_on\_B & 10 & 2453 & 0.2157031924 & 1.74234837 \\
  & B\_deposit\_on\_BW & 15 & 3293 & 0.3235547886 & 2.338994367 \\
  & WxBx\_surf\_sliced & 6 & 888 & 0.1294219154 & 0.6307400541 \\
  & B\_surface\_disp & 14 & 2016 & 0.3019844694 & 1.431950393 \\
  & surface\_disp & 30 & 2214 & 0.6471095772 & 1.572588378 \\
  & EXTD\_surface\_disp & 26 & 4770 & 0.5608283003 & 3.388096912 \\
  & deposition & 36 & 7511 & 0.7765314927 & 5.335009624 \\
\midrule
Liquid & liquid-solid & 150 & 31530 & 3.235547886 & 22.39553368 \\
\bottomrule
\end{tabular}
\label{tab:dataset-tabWB}
\end{table}

  \begin{table}[htbp]
  \centering
  \begin{tabular}{|c|c|c|c|c|}
  \hline
  Ion & Material & SY & RY & Channelling (\%) \\
  \hline
  B & alphaB \hkl(001)      & 0.054 & 0.004 & 3.27 \\
    & alphaB \hkl(010)/\hkl(100)  & 0.043 & 0.001 & 0.13 \\
    & BetaB \hkl(001)       & 0.046 & 0.003 & 1.00 \\
    & BetaB \hkl(010)/\hkl(100)   & 0.029 & 0 & 0.20 \\
    & W \hkl(100)         & 0.064 & 0.375 & 2.87 \\
    & W \hkl(110)         & 0.123 & 0.458 & 0.13 \\
    & W \hkl(111)         & 0.040 & 0.263 & 1.13 \\
    & W \hkl(112)         & 0.069 & 0.341 & 0.27 \\
  \hline
  W & alphaB \hkl(001)      & 0 & 0 & 0 \\
    & alphaB \hkl(010)/\hkl(100)  & 0 & 0 & 0.53 \\
    & BetaB \hkl(001)       & 0 & 0 & 2.00 \\
    & BetaB \hkl(010)/\hkl(100)   & 0 & 0 & 1.00 \\
    & W \hkl(100)         & 0.225 & 0.001 & 0.07 \\
    & W \hkl(110)         & 0.292 & 0 & 0.00 \\
    & W \hkl(111)         & 0.257 & 0.011 & 0.07 \\
    & W \hkl(112)         & 0.227 & 0.029 & 0.00 \\
  \hline
  \end{tabular}
  \caption{Sputtering yields (SY), reflection yields (RY) and channelling at 200 eV.}
  \label{tab:dataset-sput-200ev}
  \end{table}

  \begin{table}[htbp]
  \centering
  \begin{tabular}{|c|c|c|c|c|}
  \hline
  Ion & Material & SY & RY & Channelling (\%) \\
  \hline
  B & alphaB \hkl(001)      & 0.138 & 0.002 & 1.07 \\
    & alphaB \hkl(010)/\hkl(100)  & 0.125 & 0.003 & 0.33 \\
    & BetaB \hkl(001)       & 0.127 & 0.003 & 0.07 \\
    & BetaB \hkl(010)/\hkl(100)   & 0.103 & 0.001 & 0 \\
    & W \hkl(100)         & 0.113 & 0.244 & 4.53 \\
    & W \hkl((110)         & 0.237 & 0.358 & 0.00 \\
    & W \hkl(111)         & 0.110 & 0.223 & 3.60 \\
    & W \hkl(112)         & 0.198 & 0.351 & 0.13 \\
  \hline
  W & alphaB \hkl(001)      & 0.023 & 0 & 0.00 \\
    & alphaB \hkl(010)/\hkl(100)  & 0.019 & 0 & 0.00 \\
    & BetaB \hkl(001)       & 0.028 & 0 & 0.00 \\
    & BetaB \hkl(010)/\hkl(100)   & 0.014 & 0 & 0.20 \\
    & W \hkl(100)         & 0.740 & 0.017 & 0.33 \\
    & W \hkl(110)         & 0.791 & 0.001 & 0.33 \\
    & W \hkl(111)         & 0.631 & 0.025 & 0.53 \\
    & W \hkl(112)         & 0.611 & 0.032 & 0.20 \\
  \hline
  \end{tabular}
  \caption{Sputtering yields (SY), reflection yields (RY) and channelling at 500 eV.}
  \label{tab:dataset-sput-500ev}
  \end{table}

  \begin{table}[htbp]
  \centering
  \begin{tabular}{|c|c|c|c|c|}
  \hline
  Ion & Material & SY & RY & Channelling (\%) \\
  \hline
  B & alphaB \hkl(001)      & 0.224 & 0.004 & 0.47 \\
    & alphaB \hkl(010)/\hkl(100)  & 0.222 & 0.003 & 2.00 \\
    & BetaB \hkl(001)       & 0.242 & 0.003 & 0.13 \\
    & BetaB \hkl(010)/\hkl(100)   & 0.173 & 0.002 & 0.53 \\
    & W \hkl(100)         & 0.155 & 0.239 & 4.60 \\
    & W \hkl(110)         & 0.301 & 0.345 & 0.00 \\
    & W \hkl(111)         & 0.161 & 0.180 & 3.27 \\
    & W \hkl(112)         & 0.295 & 0.345 & 0.07 \\
  \hline
  W & alphaB \hkl(001)      & 0.113 & 0 & 0 \\
    & alphaB \hkl(010)/\hkl(10)0  & 0.095 & 0 & 0 \\
    & BetaB \hkl(001)       & 0.111 & 0 & 0 \\
    & BetaB \hkl(010)/\hkl(100)   & 0.093 & 0 & 0.07 \\
    & W \hkl(100)         & 1.583 & 0.019 & 1.27 \\
    & W \hkl((110)         & 1.485 & 0.001 & 1.13 \\
    & W \hkl(111)         & 1.225 & 0.020 & 1.53 \\
    & W \hkl(112)         & 1.293 & 0.027 & 2.00 \\
  \hline
  \end{tabular}
  \caption{Sputtering yields (SY), reflection yields (RY) and channelling at 1000 eV.}
  \label{tab:dataset-sput-1000ev}
  \end{table}

  \begin{table}[htbp]
  \centering
  \begin{tabular}{|c|c|c|c|c|c|c|}
  \hline
  Ion & Material & SY (B) & SY (W) & SY (total) & RY & Channelling (\%) \\
  \hline
  B & WB \hkl(001)       & 0.235 & 0.002 & 0.237 & 0.240 & 0.13 \\
    & WB \hkl(010) Btop   & 0.198 & 0.006 & 0.204 & 0.215 & 0.13 \\
    & WB \hkl(010) Wtop   & 0.083 & 0.024 & 0.107 & 0.321 & 0.07 \\
    & WB \hkl(100)        & 0.264 & 0.010 & 0.274 & 0.282 & 0.07 \\
  \hline 
  W & WB \hkl(001)        & 0.092 & 0.000 & 0.092 & 0 & 0.07 \\
    & WB \hkl(010) Btop   & 0.139 & 0.004 & 0.143 & 0 & 0 \\
    & WB \hkl(010) Wtop   & 0.009 & 0.110 & 0.119 & 0 & 0.07 \\
    & WB \hkl(100)        & 0.107 & 0.024 & 0.131 & 0.001 & 0.27 \\
  \hline 
  \end{tabular}
  \caption{Sputtering yields (SY), both atom-wise and total, reflection yields (RY) and channelling at 200 eV for WB surfaces.}
  \label{tab:dataset-sputWB-200ev}
  \end{table}
  
  \begin{table}[htbp]
  \centering
  \begin{tabular}{|c|c|c|c|c|c|c|}
  \hline
  Ion & Material & SY (B) & SY (W) & SY (total) & RY & Channelling (\%) \\
  \hline
  B & WB \hkl(001)        & 0.173 & 0.009 & 0.182 & 0.115 & 0.13 \\
    & WB \hkl(010) Btop   & 0.256 & 0.030 & 0.286 & 0.288 & 1.80 \\
    & WB \hkl(010) Wtop   & 0.123 & 0.055 & 0.178 & 0.298 & 2.40 \\
    & WB \hkl(100)        & 0.201 & 0.015 & 0.216 & 0.135 & 0.13 \\
  \hline
  W & WB \hkl(001)        & 0.285 & 0.057 & 0.342 & 0.003 & 0.20 \\
    & WB \hkl(010) Btop   & 0.195 & 0.063 & 0.258 & 0 & 0.20 \\
    & WB \hkl(010) Wtop   & 0.065 & 0.205 & 0.270 & 0.005 & 0.47 \\
    & WB \hkl(100)        & 0.265 & 0.091 & 0.356 & 0.013 & 0.60 \\
  \hline
  \end{tabular}
  \caption{Sputtering yields (SY), both atom-wise and total, reflection yields (RY) and channelling at 500 eV for WB surfaces.}
  \label{tab:dataset-sputWB-500ev}
  \end{table}
  
  \begin{table}[htbp]
  \centering
  \begin{tabular}{|c|c|c|c|c|c|c|}
  \hline
  Ion & Material & SY (B) & SY (W) & SY (total) & RY & Channelling (\%) \\
  \hline
  B & WB \hkl(001)        & 0.161 & 0.019 & 0.180 & 0.115 & 2.93 \\
    & WB \hkl(010) Btop   & 0.299 & 0.049 & 0.348 & 0.294 & 0.87 \\
    & WB \hkl(010) Wtop   & 0.144 & 0.085 & 0.229 & 0.319 & 1.47 \\
    & WB \hkl(100)        & 0.136 & 0.026 & 0.162 & 0.115 & 2.40 \\
  \hline
  W & WB \hkl(001)        & 0.514 & 0.149 & 0.663 & 0.011 & 1.00 \\
    & WB \hkl(010) Btop   & 0.361 & 0.142 & 0.503 & 0 & 1.13 \\
    & WB \hkl(010) Wtop   & 0.196 & 0.402 & 0.598 & 0.007 & 1.07 \\
    & WB \hkl(100)        & 0.442 & 0.204 & 0.646 & 0.015 & 3.13 \\
  \hline
  \end{tabular}
  \caption{Sputtering yields (SY), both atom-wise and total, reflection yields (RY) and channelling at 1000 eV for WB surfaces.}
  \label{tab:dataset-sputWB-1000ev}
  \end{table}